\documentstyle[amsfonts]{aipproc}

\begin{document}

\title{Stabilization not for certain and the usefulness of bounds}
\author{C. Figueira de Morisson Faria,$^*$ A. Fring$^{\dagger}$ and R. Schrader$%
^{\dagger}$}
\address{$^*$Max-Planck-Institut  f\"{u}r Physik komplexer Systeme, \\
 N\"{o}thnitzer Str. 38, D-01187 Dresden,  Germany\\
$^{\dagger}$Institut f\"ur Theoretische Physik,\\
Freie Universit\"at Berlin, Arnimallee 14, D-14195 Berlin, Germany\\} 
\maketitle

\begin{abstract}
Stabilization is still a somewhat controversial issue concerning its very
existence and also the precise conditions for its occurrence. The key
quantity to settle these questions is the ionization probability, for which
hitherto no computational method exists which is entirely agreed upon. It is
therefore very useful to provide various consistency criteria which have to
be satisfied by this quantity, whose discussion is the main objective of
this contribution. We show how the scaling behaviour of the space leads to a
symmetry in the ionization probability, which can be exploited in the
mentioned sense. Furthermore, we discuss how upper and lower bounds may be
used for the same purpose. Rather than concentrating on particular
analytical expressions we obtained elsewhere for these bounds, we focus in
our discussion on the general principles of this method. We illustrate the
precise working of this procedure, its advantages, shortcomings and range of
applicability. We show that besides constraining possible values for the
ionization probability these bounds, like the scaling behaviour, also lead
to definite statements concerning the physical outcome. The pulse shape
properties which have to be satitisfied for the existence of asymptotical
stabilization is the vanishing of the total classical momentum transfer and
the total classical displacement and not smoothly switched on and off
pulses. Alternatively we support our results by general considerations in
the Gordon-Volkov perturbation theory and explicit studies of various pulse
shapes and potentials including in particular the Coulomb- and the delta
potential.
\end{abstract}


\section*{Introduction}

There
\footnotetext{To appear in the Proceedings of the ICOMP8 (Monterey (USA), October 1999)}
is considerable interest in the high intensity regime (intensities
larger than 3.5$\times $Wcm$^{-2}$ for typical frequencies), because since
the early nineties it may be realized experimentally. The perturbative
description, which was a very successful approach in the low intensity
regime, breaks down for such high intensities. Thus, this regime constitutes
a new challenge to theorists. Comparing the status of the understanding and
clarity of the description of the two regimes one certainly observes a clear
mismatch and should probably conclude that the challenge has not been
entirely met so far. One also observes a clear imbalance between numerical
calculations and analytical descriptions.

In particular, the issue of stabilization has led to several controversies
and there are still several recent computations which are in clear
contradiction to each other. Since it is not very constructive simply to
count the numbers of numerical results which agree and those which do not%
\footnote{%
Panel discussion at this meeting.}, our investigations aim at analytical
descriptions which unravel the physical assumptions and might serve to
pinpoint possible errors.

In view of the panel discussion at this meeting the main purpose of this
contribution is to summarize our findings [1-6] and in particular explain
the working and limitations of our method in the hope to dispel a few
misunderstandings and misconceptions which have occurred.

\section*{Framework and physical properties}

We start by stating our physical assumptions. We consider an atom with
potential $V\left( \vec{x}\right) $ in the presence of a sufficiently
intense laser field, such that it may be described in the non-relativistic
regime by the time-dependent Schr\"{o}dinger equation in the dipole
approximation 
\begin{equation}
i\frac{\partial \psi (\vec{x},t)}{\partial t}=\left( -\frac{\Delta }{2}%
+V\left( \vec{x}\right) +\,\vec{x}\cdot \vec{E}\left( t\right) \right) \psi (%
\vec{x},t)=H\left( \vec{x},t\right) \psi (\vec{x},t).  \label{Schro}
\end{equation}
We will use atomic units throughout this article. We take the pulse to be of
the general form 
\begin{equation}
\vec{E}(t)=\vec{E}_{0}f(t)  \label{pulse}
\end{equation}
where $f(t)$ is assumed to be a function whose integral over $t$ is well
behaved with $f(t)=0$ unless $0\leq t\leq \tau $. This means $\tau $
constitutes the pulse duration, $f(t)$ the pulse shape function and $E_{0}$
the amplitude of the pulse, which we take to be positive without loss of
generality.

Important quantities for our discussion are the total classical momentum
transfer $\vec{b}(\tau )$, the classical displacement $\vec{c}(\tau )$ and
the classical energy transfer $a(\tau )$ defined through the relations

\begin{equation}
\vec{b}(t)=\int_{0}^{t}ds\vec{E}(s),\quad \quad \vec{c}(t)=\int_{0}^{t}ds%
\vec{b}(s),\quad \quad a\left( t\right) =\frac{1}{2}\int_{0}^{t}ds\,b^{2}%
\left( s\right) \,\,.  \label{b3}
\end{equation}
The quantity of interest, which one aims to compute, is the ionization
probability ${\cal P}\left( \varphi \right) $ defined as 
\begin{equation}
{\cal P}\left( \varphi \right) =\left\| \left( 1-P\right) U\left( \tau
,0\right) \varphi \right\| ^{2}=1-\left\| PU\left( \tau ,0\right) \varphi
\right\| ^{2}.
\end{equation}
Here $P$ denotes the orthogonal projection in the space $L^{2}({\Bbb {R}}^{3}%
{\Bbb )}$ of square integrable wave functions onto the subspace spanned by
the bound states $\varphi $ of $H\left( \vec{x},t=0\right) $, $\left\| \cdot
\right\| $ is the usual Hilbert space norm and the time evolution operator
is defined by 
\begin{equation}
U\left( t,t^{^{\prime }}\right) \equiv T[\text{Exp}(-i\int_{t^{\prime
}}^{t}H(\vec{x},s)ds)]\,\,,  \label{timeevol}
\end{equation}
with $T$ denoting the time ordering. The question one is interested in is:
How does ${\cal P}(\varphi )$ behave as a function of $E_{0}$? In particular
is it possible that ${\cal P}(\varphi )$ decreases when the field amplitude $%
E_{0}$ increases, in other words does stabilization exist? Quantitatively
this means we should find a behaviour of the form 
\begin{equation}
\frac{d{\cal P}\left( \varphi \right) (E_{0})}{dE_{0}}\leq 0\qquad \text{for}%
\quad {\cal P}\left( \varphi \right) \neq 1  \label{ppp}
\end{equation}
with $0\leq E_{0}\leq \infty $ on a finite interval for $E_{0}$. We refer to
a behaviour in (\ref{ppp}) for the equal sign as weak stabilization and for
strict inequality we call this strong stabilization.

We stress once more that this description is entirely non-relativistic. The
relativistic regime surely poses a new challenge and a full quantum field
theoretical treatment is desirable, but it should be possible to settle the
question just raised within the framework outlined above, since
stabilization is not claimed to\ be a relativistic effect. In particular it
is not clear which consequences on the physics in this regime one expects
from a description in the form of the Klein-Gordon equation\footnote{%
See contribution to the panel discussion at this meeting by F.H.M. Faisal.}.
Furthermore, appealing to a more formal description\footnote{%
See contributions to the panel discussion at this meeting by F.H.M. Faisal
and H. Reiss.} in terms of scattering matrices\footnote{%
For pulses of the form (\ref{pulse}) the scattering matrix $S=\lim_{t_{\pm
}\rightarrow \pm \infty }\exp (it_{+}H_{+})\cdot U(t_{+},t_{-})\cdot \exp
(-it_{-}H_{-})\;$and $U(\tau ,0)$ coincide in the weak sense. (see e.g. \cite
{FKS} for a more detailed discussion)} instead of the time evolution
operator $U\left( t,t^{^{\prime }}\right) $ will not shed any new light on
the question raised, unless one deals with non-trivial asymptotics.

The time-ordering in (\ref{timeevol}) poses the main obstacle for the
explicit computations of ${\cal P}\left( \varphi \right) $. To get a handle
on the issue, one can first resort to general arguments which provide
analytical expressions constraining the outcome. The least such arguments
are good for is to serve as consistency checks for results obtained by other
means. This is especially useful when one has a controversy as in the case
at hand. In addition we will demonstrate that they also allow some definite
statements and explain several types of physical behaviour without knowing
the exact expression of the quantities which describe them.

\section*{Constraints from scaling properties}

More details concerning the arguments of this section may be found in \cite
{FFS4}. Denoting by $\lambda >0$ the dilatation factor and by $\eta $ the
scaling dimension of the eigenfunction $\varphi (\vec{x}):=\psi (\vec{x}%
,t=0) $ of the Hamiltonian $H\left( \vec{x},t=0\right) $, we consider the
following scale transformations\footnote{%
More formally we could also carry out all our computations by using unitary
dilatation operators $U(\lambda )$, such that the transformation of the
eigenfunction is described by $U(\lambda )\varphi (\vec{x})=\lambda ^{\eta
}\varphi ^{\prime }(\lambda \vec{x})$ and operators ${\cal O}$ acting on $%
\varphi (\vec{x})$ transform as $U(\lambda ){\cal O}U(\lambda )^{-1}={\cal O}%
^{\prime }$.} 
\begin{equation}
\vec{x}\rightarrow \vec{x}^{\prime }=\lambda \vec{x}\quad \quad \text{%
and\quad \quad }\varphi (\vec{x})\rightarrow \varphi ^{\prime }(\vec{x}%
^{\prime })=\lambda ^{-\eta }\varphi (\vec{x})\,\,.  \label{dilatation}
\end{equation}
As the only two physical assumptions we now demand that the Hilbert space
norm, i.e. $\left\| \varphi (\vec{x})\right\| =\left\| \varphi ^{\prime }(%
\vec{x}^{\prime })\right\| $, remains invariant and that the scaling of the
wavefunction is preserved for all times. From the first assumption we deduce
immediately that the scaling dimension has to be $\eta =d/2$ with $d$ being
the dimension of the space.

The scaling behaviour (\ref{dilatation}) may usually be realized by scaling
the coupling constant. Considering for instance the wavefunction $\varphi
(x)=\sqrt{\alpha }\exp (-\alpha |x|)$ of the only bound state when the
potential in (\ref{Schro}) is taken to be the one-dimensional
delta-potential $V(x)=\alpha \delta (x)$, equation (\ref{dilatation})
imposes that the coupling constant has to scale as $\alpha \rightarrow
\alpha ^{\prime }=\lambda ^{-1}\alpha $. Choosing instead the Coulomb
potential in the form $V(\vec{x})=\alpha /r$ requires the same scaling
behaviour of the coupling constant for (\ref{dilatation}) to be valid. This
is exhibited directly by the explicit expressions of the corresponding
wavefunctions $\varphi _{nlm}(\vec{x})\sim \alpha ^{3/2}(\alpha r)^{l}\exp
(-\alpha r/n)L_{n+l}^{2l+1}(2\alpha r/n)$.

From the second assumption we conclude 
\begin{equation}
\psi (\vec{x},t)\rightarrow \psi ^{\prime }(\vec{x}^{\prime },t^{\prime
})=U^{\prime }(t^{\prime },0)\varphi ^{\prime }(\vec{x}^{\prime })=\lambda
^{-d/2}\psi (\vec{x},t)=\lambda ^{-d/2}U(t,0)\varphi (\vec{x})\,\,.
\end{equation}
Consequently this means that the time evolution operator should be an
invariant quantity under these transformations 
\begin{equation}
U(t_{1},t_{0})=T\left( e^{-i\int_{t_{0}}^{t_{1}}H(\vec{x},s)ds}\right)
\rightarrow U^{\prime }(t_{1}^{\prime },t_{0}^{\prime })=T\left(
e^{-i\int_{\lambda ^{2}t_{0}}^{\lambda ^{2}t_{1}}H^{\prime }(\vec{x}%
,s)ds}\right) =U(t_{1},t_{0})\,\,.  \label{tevol}
\end{equation}
Equation (\ref{tevol}) then suggests that the scaling of the time has to be
compensated by the scaling of the Hamiltonian in order to achieve
invariance. Scaling therefore the time as 
\begin{equation}
t\rightarrow t^{\prime }=\lambda ^{\eta _{t}}t\,\,,  \label{tscale}
\end{equation}
equation (\ref{tevol}) only holds if the Stark Hamiltonian of equation (\ref
{Schro}) scales as 
\begin{equation}
H\left( \vec{x},t\right) \rightarrow H^{\prime }\left( \vec{x}^{\prime
},t^{\prime }\right) =\lambda ^{\eta _{H}}H\left( \vec{x},t\right) \,\,\quad 
\text{with \quad }\eta _{H}=-\eta _{t}\,\,.  \label{HD}
\end{equation}
The properties (\ref{tscale}) and (\ref{HD}) could also be obtained by
demanding the invariance of the Schr\"{o}dinger equation (\ref{Schro}). The
overall scaling behaviour of $H\left( \vec{x},t\right) $ is governed by the
scaling of the Laplacian, such that we obtain the further constraint 
\begin{equation}
\eta _{H}=-2\,\,.  \label{eh}
\end{equation}
$\,$As a consequence we can read off the scaling properties of the potential
as 
\begin{equation}
V\left( \vec{x}\right) \rightarrow V^{\prime }\left( \vec{x}^{\prime
}\right) =\lambda ^{-2}V\left( \vec{x}\right) \quad \,\,\,.\,  \label{VE}
\end{equation}
Considering for instance the one-dimensional delta-potential and the Coulomb
potential in the forms specified above, equation (\ref{VE}) imposes that the
coupling constant has to scale as $\alpha \rightarrow \alpha ^{\prime
}=\lambda ^{-1}\alpha $ in both cases. This behaviour of the coupling
constant is in agreement with our earlier observations for the corresponding
wavefunctions.

\noindent We will now discuss the constraint resulting from equation (\ref
{HD}) on the scaling behaviour of the pulse. We directly observe that 
\begin{equation}
\vec{E}\left( t\right) \rightarrow \vec{E}^{\prime }\left( t^{\prime
}\right) =\lambda ^{-3}\vec{E}\left( t\right) \,\,\,\,.\quad
\end{equation}
This equation is not quite as restrictive as the one for the potential,
since in the latter case we could determine the behaviour of the coupling
whereas now a certain ambiguity remains in the sense that we can only deduce 
\begin{equation}
\vec{E}_{0}\rightarrow \vec{E}_{0}^{^{\prime }}=\lambda ^{\eta _{E_{o}}}\vec{%
E}_{0}\,,\quad f\left( t\right) \rightarrow f^{\prime }\left( t^{\prime
}\right) =\lambda ^{\eta _{f}}f\left( t\right) ,\quad \text{with }\eta
_{E_{0}}+\eta _{f}=-3\,.
\end{equation}
Thus, under the assumptions we have made, it is not possible to disentangle
the contribution coming from the scaling of the amplitude or the pulse shape
function. However, there might be pulse shape functions for which $\eta _{f}$
has to be $0$, since no suitable parameter, analogously to the coupling
constant for the potential, is available in its explicit form to achieve the
expected scaling.

Finally, we come to the scaling behaviour of the ionization probability.
Noting that the projection operator has to be a scale invariant quantity,
i.e. $P\rightarrow P^{\prime }=P$, we obtain together with (\ref{dilatation}%
) and (\ref{tevol}) that the ionization probability remains an invariant
quantity under the scaling transformation 
\begin{equation}
{\cal P}\left( \varphi \right) =\left\| \left( 1-P\right) U\left( \tau
,0\right) \varphi \right\| ^{2}\rightarrow {\cal P}^{\prime }\left( \varphi
^{\prime }\right) ={\cal P}\left( \varphi \right) .  \label{ion}
\end{equation}

We have therefore established that transforming the length scale corresponds
to a symmetry in the ionization probability ${\cal P}\left( \varphi \right) $%
. This symmetry can be exploited as a consistency check in various
approximation methods in numerical or analytical form as outlined in \cite
{FFS4}. In this sense the arguments of this section are similar in spirit to
those of the next section. Nonetheless, scaling properties may also be used
to explain directly certain types of physical behaviour, as for instance the
behaviour of ${\cal P}\left( \varphi \right) $ as a function of the coupling
constant (see \cite{FFS4}).

\section*{Constraints from bounds}

In this section we wish to comment on the method of computing bounds which
is alternative to computing ${\cal P}(\varphi )$ exactly. This means we
estimate expressions of the form

\begin{equation}
\Vert (1-P)U(\tau ,0)\varphi \Vert ^{2}\leq {\cal P}_{u}(\varphi )\quad 
\text{and}\quad \Vert PU(\tau ,0)\varphi \Vert ^{2}\leq 1-{\cal P}%
_{l}(\varphi )  \label{16}
\end{equation}
such that 
\begin{equation}
{\cal P}_{l}(\varphi )\leq {\cal P}(\varphi )\leq {\cal P}_{u}(\varphi )\,\,.
\end{equation}
{\em How does this work?} We can not go into all the technical details, but
we would like to illustrate the general principle of the computational steps
involved. First one should note that from a mathematical point of view there
are seldom general principles for deriving such inequalities, except for a
few elementary theorems (see e.g. \cite{inequ}). Therefore the steps in the
derivations very often do not always appear entirely compelling. In
mathematics, absolute inequalities, i.e. those which hold for all real
numbers, are important in analysis especially in connection with techniques
to prove convergence or error estimates, and in physics they have turned out
to be extremely powerful for instance in proving the stability of matter 
\cite{Elliot} or to establish properties of the entropy \cite{Wigner}.

The basic ingredients which are always exploited are the Minkowski and
H\"{o}lder inequalities 
\begin{equation}
\Vert \psi +\psi ^{\prime }\Vert \leq \Vert \psi \Vert +\Vert \psi ^{\prime
}\Vert ,\qquad \Vert \psi \psi ^{\prime }\Vert \leq \Vert \psi \Vert \cdot
\Vert \psi ^{\prime }\Vert ,
\end{equation}

\noindent used in the form 
\begin{eqnarray}
\Vert \psi -\psi ^{\prime }+X-X\Vert &\leq &\Vert \psi -X\Vert +\Vert X-\psi
^{\prime }\Vert , \\
\Vert \psi \,X\,X^{-1}\,\psi ^{\prime }\Vert &\leq &\Vert \psi X\Vert \cdot
\Vert X^{-1}\psi ^{\prime }\Vert \,\,\,,
\end{eqnarray}
where $\psi $ and $\psi ^{\prime }$ are meant to be formal objects. The aim
and sometimes the art of all considerations is now to choose $X$ such that
the loss in accuracy is minimized. One should resort here to as much
physical inspiration as possible, for instance if there is a conjecture or a
result from other sources which suggests a dynamics one can compare with.
There exist also more sophisticated possibilities to estimate the norm, as
for instance to relate the Hilbert space norm to different types of norms,
e.g. the operator norm\footnote{%
The operator norm is defined as $\Vert A\Vert _{op}=$sup$_{\varphi :\left\|
\varphi \right\| =1}\left\| A\varphi \right\| $.} or the Hilbert-Schmidt norm%
\footnote{%
Denoting by $\alpha _{1}\geq \alpha _{2}\geq \ldots $ the positive
eigenvalues of the operator $T=(A^{*}A)^{1/2}$ the Hilbert-Schmidt norm of
the operator $A$ is defined as $\Vert A\Vert _{H.S.}=(\sum_{n=1}^{\infty
}\alpha _{n}^{2})^{1/2}$.} 
\begin{equation}
\Vert A\psi \Vert \leq \Vert A\Vert _{op}\Vert \psi \Vert \leq \Vert A\Vert
_{H.S.}\Vert \psi \Vert \,\,.\,\,
\end{equation}

\noindent {\em Where do we start?} In fact, the starting point is identical
to the one of perturbation theory, that is the Du Hamel formula involving
the time evolution operator associated to two different Hamiltonians $%
H_{1}(t)$ and $H_{2}(t)$%
\begin{equation}
U_{1}(t,t^{^{\prime }})=U_{2}(t,t^{^{\prime }})-i\int_{t^{\prime
}}^{t}ds\,\,U_{1}\left( t,s\right) \left( H_{1}(s)-H_{2}(s)\right)
U_{2}(s,t^{^{\prime }})\,\,\,\,\,.  \label{DuH}
\end{equation}
For instance, identifying the Stark Hamiltonian in (\ref{Schro}) with $%
H_{1}(s)$, one chooses $H_{2}(s)=-\Delta /2+\,\vec{x}\cdot \vec{E}\left(
t\right) $ or $H_{2}(s)=-\Delta /2+V\left( \vec{x}\right) $ in the high- or
low intensity regime, respectively. Instead of iterating (\ref{DuH}) and
ending up with a power series in $V$ in the former or a power series in $%
E_{0}$ in the second case one inserts (\ref{DuH}) into (\ref{16}) and
commences with the estimation of the norm in the way just outlined. Most
conveniently these considerations are carried out in a different gauge, for
the high intensity regime in the Kramers-Henneberger gauge.

\noindent {\em Where do we stop?} The whole procedure may be terminated when
one arrives at expressions which may be computed explicitly.

\noindent {\em When can we apply bounds? }In general in all circumstances.
In particular problems occurring in the context of perturbative
considerations, like the convergence, are avoided completely. Especially
when the strength of the potential and the field are comparable, e.g. in the
turn-on and off region, this method is not limited in its applicability, as
is for instance the case for the Gordon-Volkov series.

\noindent {\em What can we deduce?} Ideally $P_{l}(\varphi )$ and $%
P_{u}(\varphi )$ are very close to each other, in which case we are in the
position of someone solving the problem numerically with $P_{l}(\varphi )$
and $P_{u}(\varphi )$ related to the numerical errors. If the lower bound
tends to 1 for an increasing finite realistic value of $E_{0}$ there will be
little room left for ${\cal P}\left( \varphi \right) $ to decrease and one
may deduce that stabilization is absent (see figure 9 in \cite{FFS}).
Furthermore, we can always make statements about the extreme limits. For
instance for the extreme frequency limit we obtain 
\begin{equation}
\frac{d}{dE_{0}}\left( \lim\limits_{\omega \rightarrow \infty }{\cal P}%
\left( \varphi \right) \right) =0\,\,\,\,.  \label{limo}
\end{equation}
This relates our discussion to the seminal paper on the stabilization issue
by Gavrila and Kaminski \cite{Gav}. For the extreme field amplitude limit we
found 
\begin{eqnarray}
\lim_{E_{0}\rightarrow \infty }{\cal P}\left( \varphi \right) &=&1-\left|
\left\langle \varphi ,\psi _{GV}(\tau )\right\rangle \right| ^{2}\quad \text{%
for }b(\tau )=c(\tau )=0  \label{lim} \\
\lim_{E_{0}\rightarrow \infty }{\cal P}\left( \varphi \right) &=&1\qquad
\qquad \qquad \qquad \quad \text{otherwise ,}  \label{lim2}
\end{eqnarray}
where $\psi _{GV}(\tau )=U_{GV}(\tau ,0)\varphi $ is the Gordon-Volkov wave
function. For the definition of $U_{GV}$ see (\ref{UGV}). We would like to
stress that this limit is not merely of mathematical interest\footnote{%
See contribution to the panel discussion at this meeting by F.H.M. Faisal.}.
The result (\ref{lim}) is a clear indication of weak stabilization, though
it is still desirable to find the precise onset of this behaviour. It is
also clear that as a consequence of (\ref{lim}) a value of ${\cal P}\left(
\varphi \right) $ which is equal or larger than the r.h.s. of (\ref{lim})
for {\bf any finite} and experimentally realisable value of $E_{0}$
immediately implies the existence of strong stabilization.

\noindent {\em What are the shortcomings?} For realistic values of the
parameters involved the expressions sometimes yield 
\begin{equation}
P_{l}(\varphi )=0\qquad \text{or}\qquad P_{u}(\varphi )=1  \label{01}
\end{equation}
in which case the constraint is of course not very powerful. In that
situation it simply means that we have lost too much accuracy in the
derivation for that particular parameter setting. One should note, however,
\ there is no need to give up in that situation since as is evident {\bf the
expressions for the bounds are by no means unique}. It should then be quite
clear that one can not deduce\footnote{%
As was done by J.H. Eberly at the panel discussion at this meeting.} that 
{\bf the} bound is useless if one encounters the situation (\ref{01}). Even
more such a conclusion seems very much astray in the light of \cite
{FKS,FFS,FFS3,FFS2}, where we presented numerous examples for which the
bounds are well beyond the values in (\ref{01}). Sometimes this could,
however, only be achieved for extremely short pulses. As we pointed out in 
\cite{FFS} this can be overcome at the cost of having to deal with higher
Rydberg states\footnote{%
This should not lead to the conclusion that bounds in general are
exclusively applicable to higher Rydberg states, see contribution to the
panel discussion at this meeting by M. Gavrila.}, which is a direct
consequence of the scaling behaviour outlined in the previous section.

\noindent {\em How do typical expressions look like? }In \cite{FKS} we
derived for instance the expression {\em \ } 
\begin{eqnarray}
&&{\cal \!\!\!\!\!\!\!\!\!\!\!\!\!\!\!\!P}_{l}(\varphi )=1-\Bigg\{%
\int_{0}^{\tau }\Vert (V(\vec{x}-c(t)e_{z})-V(\vec{x}))\varphi \Vert dt+ 
\nonumber \\
&&\!\!\!\!\!\!\!\!\!\!\!\!\!\!\!\!\!\!\!\frac{2}{2E+b(\tau )^{2}}\Vert (V(%
\vec{x}-c(\tau )e_{z})-V(\vec{x}))\varphi \Vert +\frac{2|b(\tau )|}{%
2E+b(\tau )^{2}}\Vert p_{z}\varphi \Vert \Bigg\}^{2}  \label{lb}
\end{eqnarray}
\smallskip for a lower bound. For given potentials and pulse shapes terms
involved in (\ref{lb}) may be computed at ease. As stated in \cite{FKS}, it
is important to pay attention to the fact that (\ref{lb}) is derived for the
condition $b(\tau )^{2}/2>-E\equiv $ binding energy\footnote{%
During the panel discussion at this meeting J.H. Eberly exhibited a plot of
our result for ${\cal P}_{l}(\varphi )$ involving a pulse which did not
satisfy this condition. As he confirmed to a question from the audience his
pulse satisfied $b(\tau )=0$. The conclusions drawn by J.H. Eberly
concerning the usefulness of bounds based on this plot are therefore
meaningless. (See also footnote 9.)}. Such restrictions which at first
emerge as technical requirements in the derivations usually indicate at some
physical implications. In this case it points at the different physical
situation we encounter when the total momentum transfer is vanishing (see
also (\ref{lim})).

\noindent {\em What still needs to be done? }Probably it is unrealistic to
expect to find a bound which is universally applicable and restrictive at
the same time, rather one should optimize the bounds for particular
situations. For instance it would be highly desirable to find more powerful
bounds for the situations $b(\tau )=0,c(\tau )\neq 0$ and $b(\tau )=c(\tau
)=0$. For the latter case we expect in hindsight from (\ref{lim}) that the
loss in the estimations may be minimized if in (\ref{DuH}) we chose to
compare the Stark Hamiltonian with the free Hamiltonian $-\Delta /2$ instead
of $H=-\Delta /2+\,\vec{x}\cdot \vec{E}\left( t\right) $ as was done in \cite
{FKS}.

\section*{Importance of pulse shapes}

From our previous discussion it is evident that the physical outcome differs
for different pulse shapes. However, the fact that a pulse is adiabatically
switched on or off is not very important, rather the precise values of $%
b(\tau )$ and $c(\tau )$ are the determining quantities. In particular the
case 
\begin{equation}
b(\tau )=c(\tau )=0  \label{bc0}
\end{equation}
is very special, since then asymptotically weak stabilization is certain to
exist. An adiabatically switched on or off pulse sometimes satisfies (\ref
{bc0}), but this condition is by no means identical to it. We found no
evidence for stabilization for an adiabatically switched on field when {\bf %
\ }$b(\tau )\neq 0$. To our knowledge the importance of (\ref{bc0}) was
first pointed out by Grobe and Fedorov \cite{GroF}, using intuitive
arguments, who employed a trapezoidal enveloping function with symmetrical
turn-on and turn-off time $T$, which has the nice feature that for $T$ and $%
\tau $ being integer cycles $b(\tau )=c(\tau )=0$ and for $T$ half $\tau $
being integer cycles $b(\tau )=0$, $c(\tau )\neq 0$. Thereafter, this
observation seems to have been widely ignored in the literature since many
authors still employ pulses which do not have this property, trading (\ref
{bc0}) for the condition of an adiabatic smooth turn-on or/and turn-off%
\footnote{%
As may be supported by numerous publications, this observation appears not
to have become common knowledge as claimed by M. Gavrila in the introduction
to the panel discussion at this meeting.}. For instance a sine-squared
switch on and off with $T$ and $\tau $ being integer cycles has $b(\tau )=0$%
, $c(\tau )\neq 0$, an entire sine-squared envelope for $\tau $ being
integer cycles satisfies $b(\tau )=0$, $c(\tau )\neq 0$. Using gau\ss ian
envelopes or gau\ss ian switch on and no switch off usually yields $b(\tau
)\neq 0$, $c(\tau )\neq 0$. A pulse which has the nice features that it
allows a theoretical investigation of all possible cases for the values of $%
b(\tau )$ and $c(\tau )$ is the triple $\delta $-kick in the form 
\begin{equation}
f(t)=\delta (t)+\beta _{1}\delta (t-\tau /2)+\beta _{2}\delta (t-\tau )\,\,,
\end{equation}
which we employed in \cite{FFS2}. This pulse obviously satisfies 
\begin{equation}
b(\tau )=E_{0}(1+\beta _{1}+\beta _{2}/2)\qquad \text{and\qquad }c(\tau
)=E_{0}(1+\beta _{1}/2)
\end{equation}
such that by tuning the constants $\beta _{1},\beta _{2}$ we may realise any
desired value of $b(\tau )$ and $c(\tau )$.

\noindent How do real pulses look like\footnote{%
We acknowledge that the following argument was initiated, though not agreed
upon in this form, by an e-mail communication with H.G. Muller.}? The
quantity which is experimentally accessible is the Fourier transform of the
pulse (\ref{pulse}) 
\begin{equation}
\widetilde{E}(\omega )=\int_{-\infty }^{\infty }E(t)e^{i\omega
t}dt\,=\sum_{n=0}^{\infty }\alpha _{n}\omega ^{n}\ .
\end{equation}

\noindent with $\alpha _{n}$ being constants. For finite pulses this
quantity coincides with the total momentum transfer for vanishing frequency $%
\omega $ 
\begin{equation}
\widetilde{E}(\omega =0)=\int_{-\infty }^{\infty }E(t)dt=\int_{0}^{\tau
}E(t)dt=b(\tau )\,\,.
\end{equation}
Provided that $\alpha _{0}=b(\tau )=0$, the Fourier transform of the
momentum transfer 
\begin{equation}
\widetilde{b}(\omega )=\int_{-\infty }^{\infty }b(t)e^{i\omega t}dt\,\,
\end{equation}
is on the other hand related to the total displacement for vanishing
frequency 
\begin{equation}
\widetilde{b}(\omega =0)=\int_{-\infty }^{\infty }b(t)dt=\int_{0}^{\tau
}b(t)dt=c(\tau )\quad
\end{equation}
such that 
\begin{equation}
\widetilde{E}(\omega )=b(t)e^{i\omega t}|_{-\infty }^{\infty }-i\omega 
\widetilde{b}(\omega )\sim -i\omega c(\tau )+{\cal O}(\omega ^{2})\,\,.
\end{equation}
This means that when the experimental outcome is

\begin{equation}
\widetilde{E}(\omega )=\alpha _{2}\omega ^{2}+\alpha _{3}\omega ^{3}+\alpha
_{4}\omega ^{4}+\ldots
\end{equation}
the total momentum transfer and the total displacement are zero.
Experimentally, the observed fall off is expected to be even stronger \cite
{Muell}.

\section*{Comparison with GV-perturbation theory}

It is instructive to compare our findings with other standard methods as for
instance the Gordon-Volkov (GV) perturbation theory. Using now in (\ref{DuH}%
) for $H_{2}$ the Hamiltonian just involving the field and the free particle
Hamiltonian in the Kramers-Henneberger frame subsequent iteration yields\ 
\begin{eqnarray}
U_{1}(t,t^{^{\prime }}) &=&U_{GV}(t,t^{^{\prime }})-i\int_{t^{\prime
}}^{t}ds\,\,U_{GV}\left( t,s\right) VU_{GV}(s,t^{^{\prime }})  \nonumber \\
&&\!\!\!\!\!\!\!\!\!\!\!\!-\int_{t^{\prime }}^{t}ds\int_{s}^{t}ds^{\prime
}\,\,U_{GV}\left( t,s^{\prime }\right) VU_{GV}(s^{\prime
},s)VU_{GV}(s,t^{^{\prime }})+\ldots
\end{eqnarray}

\noindent where $U_{GV}$ corresponds to the free-particle evolution operator
in the KH frame 
\begin{equation}
U_{GV}(t,t^{^{\prime
}})=e^{-ia(t)}e^{-ib(t)z}e^{ic(t)p_{z}}e^{-i(t-t^{\prime })\frac{p^{2}}{2}%
}e^{-ic(t^{\prime })p_{z}}e^{ib(t^{\prime })z}e^{ia(t^{\prime })}.
\label{UGV}
\end{equation}
As was explained in \cite{FFS3} we may use these expressions together with
the Riemann-Lebesgue theorem in order to obtain the extreme frequency and
intensity limit, finding (\ref{limo}), (\ref{lim}) and (\ref{lim2}). For
these arguments to be valid we have to assume that the Gordon-Volkov series
makes sense, so in particular we have to assume its convergence.

The latter assumption may be made more rigorous when considering the
one-dimensional delta potential $V\left( x\right) =-\alpha \,\delta \left(
x\right) $ which is well known to possess only one bound state. In that case
the problem of computing ionization probabilities is reduced to \smallskip
the evaluation of 
\begin{equation}
{\cal P}\left( \varphi \right) =1-\left| \left\langle \varphi ,\psi
_{GV}(\tau )\right\rangle +\left\langle \varphi ,\Psi (\tau )\right\rangle
\right| ^{2}\quad
\end{equation}
\medskip with

\begin{equation}
\left\langle \varphi ,\psi _{GV}(\tau )\right\rangle =\frac{2}{\pi }%
e^{-ia(\tau )}\int_{-\infty }^{\infty }dp\frac{\exp \left( -i\tau \alpha ^{2}%
\frac{p^{2}}{2}-ic\left( \tau \right) \alpha p\right) }{\left( 1+\left(
p+b\left( \tau \right) /\alpha \right) ^{2}\right) \left( 1+p^{2}\right) }
\end{equation}
\medskip 
\begin{equation}
\left\langle \varphi ,\Psi (\tau )\right\rangle =ie^{-ia(\tau )}\sqrt{\frac{%
\alpha ^{5}}{2\pi ^{3}}}\int_{0}^{\tau }\int_{-\infty }^{\infty }\psi
_{I}\left( s\right) \frac{e^{i(c(\tau )-c(s))p}e^{-\frac{i}{2}p^{2}(\tau
-s)}dsdp}{\left( \alpha ^{2}+(p+b(\tau ))^{2}\right) }\,\,.
\end{equation}
\medskip Here the only unknown is the function $\psi _{I}\left( t\right) $
which can be obtained as a solution of the Volterra equation 
\begin{equation}
\psi _{I}\left( t\right) =\int_{-\infty }^{\infty }dp\,\psi _{GV}\left(
p,t\right) +\alpha \sqrt{\frac{i}{2\pi }}\int_{0}^{t}ds\psi _{I}\left(
s\right) \frac{e^{i\frac{(c(t)-c(s))^{2}}{2(t-s)}}}{\sqrt{t-s}}\,\,.
\end{equation}
Iteration of this equation is a well controllable procedure and in \cite
{FFS2} we found that the series converges for all values of $\alpha $. The
results obtained from the analysis of this equation match the results
obtained from bounds.

\section*{Conclusions}

The main outcome of our investigations is that the {\bf classical momentum
transfer} and {\bf displacement} caused by a laser pulse on an electron are
the essential parameters determining the existence of weak asymptotic
stabilization. In fact, we obtained evidence for stabilization only for
pulses for which these two quantities vanish at the end of the pulse, i.e.,
with $b(\tau )=0$ and $c(\tau )=0$.

Using purely analytical methods, we have shown that, for a wide range of
potentials, namely Kato and one- and three-dimensional delta potentials, we
always have $\lim_{E_{0}\rightarrow \infty }{\cal P}(\psi )=1$ unless $%
b(\tau )=0$ and $c(\tau )=0$, in which case the ionization probability tends
to the lowest order in GV-perturbation theory, which corresponds simply to
the free particle Green's function (\ref{UGV}). Furthermore, for infinite
frequencies, the high-frequency condition of \cite{Gav} is a way to obtain $%
b(t)=0$ and $c(t)=0$ for {\it all} times.

Clearly, smooth pulses in general do not necessarily fullfil the above
conditions, and therefore will not provide a mechanism for stabilization,
but just prolong the onset of ionization. In fact, we have observed no
stabilization for adiabatically switched on and off pulses of several
shapes, for which analytic expressions for lower bounds of ionization
probabilities lead to {\bf conclusive} statements concerning the existence
or absence of stabilization.

Therefore, as an overall conclusion: {\bf Bounds are useful indeed, also in
the context of high intensity laser physics!}

\end{document}